\documentclass[nohyper,12pt,letterpaper]{JHEP3}
\usepackage{youngtab, epsfig}

\author{Rajsekhar Bhattacharyya$^{2}$ and Robert de Mello Koch$^{1,2}$,\\
\qquad \\
Department of Physics and Centre for Theoretical Physics$^{1}$,\\ 
University of the Witwatersrand,\\ 
Wits, 2050,\\ 
South Africa\\
\qquad\\
Stellenbosch Institute for Advanced Studies,$^{2}$\\
Stellenbosch,\\
South Africa\\
\qquad\\
E-mail: \email{rbhattac@sun.ac.za, robert@neo.phys.wits.ac.za}}

\abstract{
It is well known that a D-string ending on a D3, D5 or D7 brane is described 
in terms of a non-commutative fuzzy funnel geometry. In this article, we give
a numerical study of the fluctuations about this leading geometry. This allows 
us to investigate issues related to the stability and moduli space of these 
solutions. We comment on the comparison to the linearized fluctuations in supergravity.}

\preprint{}

\title{Fluctuating Fuzzy Funnels}


\def \Tr{\mbox{Tr\,}}

\begin{document}

\section{Introduction}
Fuzzy spheres are fascinating examples of non-commutative geometry. 
They have obvious application in string theory, where they arise as 
solutions to matrix brane actions\cite{const}\cite{const2}\cite{kabat}\cite{cook}\cite{recent}, 
and may also play a role in a spacetime explanation of the stringy 
exclusion principle\cite{sep}. In this article we will be interested
in describing small fluctuations about the so-called ``fuzzy funnel" 
solutions describing D-strings ending on D3\cite{const}, D5\cite{const2} 
or D7\cite{cook} branes. The funnels have the topology 
$R\times S^k$ with $S^k$ a fuzzy sphere. Although all of the funnels that we consider
are time independent, the equations for the funnel also admit interesting 
time dependent solutions\cite{tdep}.

An important difference between a point particle and an extended object is the
fact that an extended object has a set of low energy excitations which arise
from small vibrations about their equilibrium configuration. These modes have
already played an important role in the study of black holes in string theory
where they account for both the microscopic entropy\cite{15} and the low energy Hawking
radiation of the black holes\cite{16}. There are a number of reasons why a study 
of the spectrum of small fluctuations of the fuzzy funnels
is interesting. First, the prescence of tachyonic modes signals an instability
in the solution. The fuzzy funnel relevant for D-strings ending on D7-branes
has been given a description from both the point of view of the D-strings and 
of the D7-branes\cite{cook}. It turns out that the two descriptions are in
good agreement, at leading order in $N$. From the point of view of the D7 
brane description, the funnel
is described by a 7+1 dimensional non-Abelian Born-Infeld theory with a specific
non-zero field strength switched on. It appears\cite{cook} that it is possible to 
lower the energy of the funnel solution, by changing this field strength. Thus
there are good reasons to question the stability of the D1$\perp$D7 funnel. Our
analysis shows that the D1$\perp$D7 system is perturbatively stable.
Second, if there is a family of funnel solutions this will show up as
zero modes in the small fluctuation spectrum. Third, the geometry of the fuzzy funnel
involves a fuzzy sphere. For the fuzzy $S^4$ and $S^6$ there are new degrees of
freedom\cite{sanjaye}\cite{ho} that have no counterpart in the commutative $S^4$ and 
$S^6$. These degrees of freedom are most important at finite $N$ corresponding to 
situations in which quantum gravity corrections can not be neglected. Understanding the 
role of these new degrees of freedom is an important problem. By studying the small 
fluctuations of the funnel, we will gain some insight into these degrees of freedom.

The fluctuations with which we concern ourselves arise from the motion of the branes
in spacetime. Consequently we do not consider the fermions or the gauge field that
resides on the branes worldvolume.
To carry out the small fluctuation analysis, we need a basis in which to expand
our fluctuations. If we were dealing with the geometry of the usual commutative 
sphere, we would simply use the spherical harmonics. Since we deal with the fuzzy
sphere we will have to use a deformed algebra, where the deformation allows for
a cut off on the angular momentum. We will see that the usual approach to 
constructing this basis for the fluctuations is frustrated by the non-commuting 
nature of the coordinates. The natural guess - to take traceless symmetric 
combinations of the underlying coordinates - does not lead to a closed algebra.
Fortunately, Ramgoolam\cite{sanjaye} has found an elegant solution to this problem.
For the $S^{2n}$, instead of simply restricting to the traceless symmetric 
representations, one considers all possible representations of $SO(2n+1)$, which
may conveniently be labelled by Young Diagrams. The cut off on angular momentum 
is implemented as a cut off on the number of columns of the Young diagrams.
Since there are only a finite number of independent basis functions in the deformed
algebra the problem lends itself to a numerical study. In this article we provide
this numerical analysis of the small fluctuations for both the D1$\perp$D5
and the D1$\perp$D7 systems.

In \cite{const2} small fluctuations of the D1$\perp$D5 system were studied.
The funnel analysis was performed where the funnel is narrow and hence it
should be reliably described in terms of $N$ D-strings. These results did
not match results obtained for a test-string in the background of $n$ D5
branes. Since both the D-string and the test string descriptions should be
valid for large $\sigma$ this disagreement is rather disturbing. One of our
goals in this article is to resolve this puzzle. Using a Newtonian approximation
we are able to produce a model which uses the full funnel geometry as opposed
to \cite{const2} where the funnel was effectively replaced by $n$ D5 branes.
We then find that the dynamics of the fluctuations depends sensitively on the 
size of the fluctuation. Further, we show that it is possible to reproduce
the funnel results using our Newtonian toy model.

In the next section we briefly review the fuzzy funnel solutions for both
the D1$\perp$D5 and the D1$\perp$D7 systems. In section 3 we discuss in detail 
how a basis for the fluctuations can be constructed, allowing for a numerical 
solution of the small fluctuations problem. In section 4, we describe our
formulation of the small fluctuations problem and in section 5 we give the 
results of this analysis. The analysis of sections 4 and 5 is only valid for
small field values. In section 6 we compare our results to a gravity
analysis. In section 7 we present our conclusions and indicate some directions
in which this work can be extended.

\section{Review of the Fuzzy Funnel Solutions}

As mentioned above, the fuzzy funnel solutions are relevant for the description of 
D-strings ending on either D3, D5 or D7 branes. The funnels have the topology 
$R\times S^k$ with $S^k$ a fuzzy sphere with $k=2$ for the D1$\perp$D3 system,
$k=4$ for the D1$\perp$D5 system and $k=6$ for the D1$\perp$D7 system. We do not concentrate 
much on the D1$\perp$D3 system in this article. The fuzzy funnels can be described 
using either a worldvolume description of the strings, or a worldvolume description
of the branes. The two descriptions are in perfect agreement. We will focus on the
description in terms of the D-strings. 

The low energy effective action for $N$ D-strings is given by the non-Abelian 
Born-Infeld action\cite{Myers},\cite{Tsyetlin}

\begin{equation}
S=-T_1\int d^2\sigma STr\sqrt{-\det\left[
\matrix{\eta_{ab} &\lambda\partial_a\Phi^j\cr 
-\lambda\partial_b\Phi^i &Q^{ij}}
\right]}\equiv -T_1\int d^2\sigma STr\sqrt{-\det M}\quad ,
\label{Nabi}
\end{equation}

\noindent
where

$$ Q^{ij} = \delta^{ij}+i\lambda\big[\Phi^i ,\Phi^j\big],\qquad
\lambda =2\pi l_s^2.$$

\noindent
The symmetrized trace prescription \cite{Tseyt} (indicated by $STr$ in the above action)
instructs us to symmetrize over all
permutations of $\partial_a\Phi^i$ and $\big[\Phi^i ,\Phi^j\big]$ within the trace over
the gauge group indices, after expanding the square root. We are using static gauge so
that the worldsheet coordinates are identified with spacetime coordinates as
$\tau =x^0$ and $\sigma =x^9$. The transverse coordinates are now the non-Abelian
scalars $\Phi^i$, $i=1,...,8$. These scalars are $N\times N$ matrices transforming in the 
adjoint representation of the $U(N)$ gauge symmetry of the worldsheet theory of the D1s. 

The lowest order equation of motion is

$$(-\partial_\tau^2+\partial_\sigma^2)\Phi^i=\big[\Phi^j,\big[\Phi^j,\Phi^i\big]\big].$$

\noindent
This equation of motion is valid for small $\Phi^j$, where the funnel is accurately
described by a collection of D1s.
We are interested in studying the fluctuations $\delta\Phi^i$ about
the leading configuration $\Phi_0^i$

$$\Phi^i =\Phi_0^i +\delta\Phi^i .$$

\noindent
The linearized equation for the fluctuation following from this lowest 
order equation of motion is

$$(-\partial_\tau^2+\partial_\sigma^2)\delta\Phi^i
=\big[\delta\Phi^j,\big[\Phi^j_0,\Phi^i_0\big]\big]
+\big[\Phi^j_0,\big[\delta\Phi^j,\Phi^i_0\big]\big]
+\big[\Phi^j_0,\big[\Phi^j_0,\delta\Phi^i\big]\big].$$

\noindent
It is this last equation that we analyze in detail in sections 4 and 5. 

In the remainder of this section, we give the leading configurations $\Phi^i_0$.
We start by briefly reviewing the construction of fuzzy spheres, and then discuss
the funnels themselves. In the next section, we will provide a basis\cite{sanjaye} 
in terms of which the fluctuations $\delta\Phi^i$ can be expanded.

\subsection{Fuzzy Spheres}
To construct the fuzzy $S^{2m}$ sphere, we need to solve the equation

\begin{equation}
\sum_{i=1}^{2m+1} X^iX^i =c{\bf 1},
\label{FSphere}
\end{equation}

\noindent
with $X^i$ a matrix, ${\bf 1}$ the identity matrix and $c$ a constant.
There is a simple construction of the matrices $X^i$, in terms of the 
Clifford algebra

$$\{\Gamma^i,\Gamma^j\}=2\delta^{ij},\qquad i,j=1,2,..., 2m+1 .$$

\noindent
The $n$-fold tensor product of $V$ (the space on which the $\Gamma^i$ matrices act) 
is written as $V^{\otimes n}$. The $X^i$ are now obtained by taking

$$ X^i =\left( \Gamma^i \otimes 1\otimes ...\otimes 1 +
1 \otimes \Gamma^i \otimes ...\otimes 1 +...+
1 \otimes 1\otimes ...\otimes \Gamma^i\right)_{st}.$$

\noindent
The subscript $st$ is to indicate that the above $X^i$ are to be restricted
to the completely symmetric tensor product space for $2m=2,4$, and symmetric 
and traceless tensor product space for $2m=6$. Clearly, the $X^i$ act in 
$V^{\otimes n}$. To prove that the above $X^i$ 
do indeed provide coordinates for the fuzzy sphere, one shows that 
$\sum_{i=1}^{2m+1} X^i X^i$ commutes with the generators of $SO(2m+1)$; then by
Schur's lemma, it follows that $\sum_{i=1}^{2m+1} X^i X^i$ is proportional to
the identity.

For properties of the matrices $X^i$ in the case of the fuzzy $S^4$, see \cite{kabat}
and in the case of the fuzzy $S^6$, see \cite{kimura}

\subsection{D1/D5 System}
We use $N$ to denote the number of D-strings in the funnel and $n$ to denote the number of
D5-branes in the funnel. Five of the eight scalars are excited

$$\Phi_0^i=R(\sigma )X^i,\qquad i=1,2,3,4,5,$$

\noindent
where the function $R(\sigma )$ is, for small $R$,

$$R(\sigma )={1\over 2\sqrt{2} \sigma }.$$

\noindent
In this case, the matrices $X^i$ have dimension

$$ N={1\over 6}(n+1)(n+2)(n+3),$$

\noindent
and $c=n(n+4)$.

\subsection{D1/D7 System}
We use $N$ to denote the number of D-strings in the funnel and $n$ to denote the number of
D7-branes in the funnel. Seven of the eight scalars are excited

$$\Phi_0^i=R(\sigma )X^i,\qquad i=1,2,...,7,$$

\noindent
where the function $R(\sigma )$ is, for small $R$,

$$ R(\sigma)={1\over 2\sqrt{3}\sigma }.$$

\noindent
In this case, the matrices $X^i$ have dimension

$$ N={1\over 360}(n+1)(n+2)(n+3)^2(n+4)(n+5),$$

\noindent
and $c=n(n+6)$.

\section{Parametrizing the Fluctuations}

To study the geometry of $S^2$ embedded in $R^3$, we can study the algebra of smooth 
functions over $S^2$. This algebra can be taken as a commutative polynomial algebra, 
i.e the polynomial ring $R[x^1, x^2, x^3]$ together with the constraint 

\begin{equation}
(x^1)^2 + (x^2)^2 + (x^3)^2 =c.
\label{constraint}
\end{equation}

\noindent
In other words we can study $R[x^1, x^2, x^3]/I$ where $I$ is the ideal generated by 
the polynomial $(x^1)^2+ (x^2)^2 + (x^3)^2-c$.

It is then natural to study the geometry of the fuzzy $S^2$ by studying the algebra of 
functions over it. For some fixed positive integer $n$ we replace $x^1, x^2, x^3$ by three 
$n\times n$ matrices $X^1, X^2, X^3$, which generate the $SU(2)$ Lie algebra 

$$[X^i, X^j]=iKC^{ijk}X^k ,$$ 

\noindent
where $C^{ijk}=\epsilon^{ijk}$ and where $K$ depends on the integer $n$. We still need 
to impose the analog of the constraint (\ref{constraint}). It turns out that there is
a one-to-one correspondence between the constraints in the case of the commutative $S^2$
and the constraints in the case of the fuzzy $S^2$. This one-to-one correspondence between
the constraints can be traced back to the fact that the $X^i$'s form a closed Lie algebra 
among themselves. Thus, the basis elements for the non-commutative algebra over the fuzzy 
$S^2$, can be obtained by replacing $x^i\to X^i$ in the basis elements of the algebra 
for the commutative $S^2$. Note however that we have an upper bound on the number of basis 
elements in the algebra over the fuzzy $S^2$, depending on the dimension of the representation 
of the $SU(2)$ Lie algebra.

For the case of $S^4$ and fuzzy $S^4$ or $S^6$ and fuzzy $S^6$ things are not so simple. 
We discuss in detail the case of the fuzzy $S^4$. For the geometry of the commutative $S^4$, 
the algebra of functions is given by $R[x^1, x^2, x^3, x^4, x^5 ]/I'$ where $I'$ is the ideal
generated by the polynomial $(x^1)^2+(x^2)^2+(x^3)^2+(x^4)^2+(x^5)^2-c$. For the fuzzy $S^4$, 
we replace $x^i$'s by the $X^i$'s given above. The $X^i$'s do not form a closed Lie algebra. 
They can be extended to the $SO(5,1)$ Lie algebra by including the new generators 

$$X^{ij}={1\over 2}\big[X^i,X^j\big].$$

\noindent 
The need to include these new elements implies that there are some constraints that do not
follow from the two sided ideal generated by 

$$(X^1)^2 + (X^2)^2 + (X^3)^2+ (X^4)^2 + (X^5)^2-cI.$$

\noindent
To illustrate this point, note that the relation\cite{const2}

$$ X^{ij} X^j - X^j X^{ji}=0,$$ 

\noindent
follows from

\begin{eqnarray}
&{1\over 2}&X^i\left(
(X^1)^2 + (X^2)^2 + (X^3)^2 + (X^4)^2 + (X^5)^2-cI\right)\nonumber\\
-&{1\over 2}&\left((X^1)^2 + (X^2)^2 + (X^3)^2 + (X^4)^2 + (X^5)^2-cI\right)X^i
\nonumber
\end{eqnarray}

\noindent
so that it can be written as an element of the two sided ideal; in contrast, the relation 

$$ X^{ij} X^j-4X^i=0,$$ 

\noindent
only follows after using the $SO(5,1)$ Lie algebra. This spoils any possible connection 
between the basis functions in the commutative and non-commutative cases. Thus, finding 
a suitable basis in which to expand the fluctuations of the fuzzy $S^4$ and fuzzy $S^6$ 
can not simply be obtained by replacing $x^i\to X^i$ in the basis for the commutative 
$S^4$ and $S^6$. 

Fortunately, Ramgoolam\cite{sanjaye} has given an elegant solution to this problem. In the
remainder of this section, we describe the basis provided by\cite{sanjaye} for both the fuzzy
$S^4$ (which is relevant for the D1$\perp$D5 system) and for the fuzzy $S^6$ (which is relevant
for the D1$\perp$D7 system). In particular, we will describe our algorithm for generating
the basis that was used in the numerical computation of the spectrum.

\subsection{D1/D5 System}

In this section we will describe an $SO(5)$ covariant basis for the $N\times N$ matrices
$\delta\Phi^i$. Elements of this basis are labelled by 
irreducible representations of $SO(5)$ which are themselves associated
with a unique Young diagram. The Young diagrams labelling the representations have at most
two rows. To translate a specific Young diagram into a basis element, we 
associate a set of indices to the Young diagram. For example, consider
the representation associated with the Young diagram with two boxes in the first row 
and one box in the second row. A suitable set of indices would be

$$\young(kl,m).$$

\noindent
Each index is associated with an $X^i$; for the above example, the basis element is built from
$X^k$, $X^l$ and $X^m$. The shape of the Young diagram gives a rule for how the $X^i$s should 
be combined: First we symmetrize over indices in the same row

$$ X^k X^l X^m\to (X^k X^l +X^l X^k)X^m = X^k X^l X^m +X^l X^k X^m,$$

\noindent
and then antisymmetrize over indices in the same column

$$ X^k X^l X^m +X^l X^k X^m \to X^k X^l X^m -X^m X^l X^k +X^l X^k X^m - X^l X^m X^k .$$

\noindent
Numerically, it is straightforward to implement this symmetrization and then the
antisymmetrization. There is an additional complication because we deal with representations 
of the orthogonal group: the above state can be written as

$$ X^k X^l X^m -X^m X^l X^k +X^l X^k X^m - X^l X^m X^k \equiv C_{abc}X^a X^b X^c.$$

\noindent
To get a state which belongs to the advertised irreducible representation, we need to
restrict ourselves not only to tensors $C_{abc}$ with the correct symmetrization and
antisymmetrization properties, but also, tensors that are traceless. Numerically we 
implemented this traceless constraint by subtracting out any projections our basis function
had on irreducible representations corresponding to Young diagrams containing a smaller
number of boxes. The cut off on angular momentum becomes a cut off on the number of columns 
in the Young diagrams. One can show\cite{sanjaye} that the number of independent states
in the irreducible representations included by this prescription is precisely equal to

$$ N^2 ={(n+1)^2(n+2)^2 (n+3)^2\over 36},$$

\noindent
and hence that we do indeed obtain a complete basis in which to express the $\delta\Phi^i$.

Numerically, we performed an exhaustive search over all possible placements of indices in 
the allowed Young diagrams. Of course, many of the operators constructed in this way were
identically zero. Further, not all of the operators constructed were independent, due to
the tracelessness constraint. Only the independent operators were kept. Our numerical results
show clearly that one does indeed construct a complete orthonormal basis using the procedure
outlined in \cite{sanjaye}.

\subsection{D1/D7 System}

In this case, one sums over irreducible representations of $SO(7)$ which can be labelled 
by Young diagrams with three rows. Again the cut off is implemented on the columns of the
Young diagram. In this case, one expects a total of\cite{sanjaye}

$$ N^2={(n+1)^2 (n+2)^2 (n+3)^4 (n+4)^2 (n+5)^2\over 129600},$$

\noindent
basis functions. Again, our numerical results are in perfect agreement with this 
expectation.

\section{Problem Formulation}

We are now ready to compute the spectrum of the small fluctuations. As discussed in
section 2, we consider the lowest order equation of motion. Thus our results will only
be valid for small $\Phi^i_0$, corresponding to the large $\sigma$ region of both
funnels. In this limit both funnels look like a collection of D-strings.

The problem of computing the small fluctuations essentially reduces to solving the
eigenvalue problem for the operator $L$ defined by ($i=1,...,8$)

$$Lv^{(C)}(i)=[v^{(C)}(j),[\phi^{j},\phi^{i}]]+
[\phi^{j},[v^{(C)}(j),\phi^{i}]]+ [\phi^{j},[\phi^{j},[v^{(C)}(i)]]
=\lambda(C) v^{(C)}(i).$$ 

\noindent
The fluctuation is labelled by $C$. The index $i$ labels the components of a particular
fluctuation. Thus $v^{(C)}(i)$, fluctuation $C$, is to 
be viewed as a vector with 8 components; each
component is an $N\times N$ matrix. Denote the basis elements described in the previous 
section by $\beta^A$, with $A=1,...,N^2$. Further, introduce a basis for the fluctuations, 
given by $|C\rangle=|\{A,i\}\rangle $. All of the components of the vector 
$|\{A,i\}\rangle $ are the $N\times N$ 
zero matrix, except the i$^{th}$ component, which is $\beta^A$. Thanks to the fact that
$L$ has a linear action on $v^{(C)}$, the eigenvectors of $L$ can be expressed as a linear
combination of the $|\{A,i\}\rangle$. Writing

$$|\{A,i\}\rangle =\left[\matrix{\beta^A\delta^{i1}\cr \beta^A\delta^{i2}\cr 
:\cr \beta^A\delta^{i8}}\right],$$

\noindent
we define the inner product

$$\langle \{A,i\}|\{A',i'\}\rangle\equiv \delta^{ii'}\Tr (\beta^{A\dagger}\beta^{A'}).$$

\noindent
We will normalize our states so that $\langle \{B,j\}|\{A,i\}\rangle=\delta^{AB}\delta^{ij}$.
Using the completeness relation

$$\sum_{A,i}|\{A,i\}\rangle\langle \{A,i\}|={\bf 1},$$

\noindent
where ${\bf 1}$ is the $8N^2\times 8N^2$ identity matrix, we can write

$$ L|\{A,i\}\rangle =\sum_{B,j}\langle \{B,j\}|L|\{A,i\}\rangle |\{B,j\}\rangle,$$

\noindent
where $\langle \{B,j\}|L|\{A,i\}\rangle$ is an $8N^2\times 8N^2$ matrix. In terms
of the eigenvectors of this matrix

$$\sum_{A,i}\langle \{B,j\}|L|\{A,i\}\rangle u^{(C)}(A,i)= \lambda (C)u^{(C)}(B,j),$$

\noindent
the eigenvectors of $L$ can be written as

$$ v^{(C)}=\sum_{A,i}u^{(C)}(A,i)|\{A,i\}\rangle .$$

\noindent
Our approach entails computing the matrix $\langle B,j|L|A,i\rangle$ and diagonalizing
it numerically to obtain $\lambda (C)$. After making the ansatz

$$\delta\Phi^i = f^{(C)}(\sigma ,\tau )v^{(C)}(i),$$

\noindent
the lowest order equation of motion then reduces to

$$\left(-\partial_\tau^2+\partial_\sigma^2 -\lambda(C)R^2(\sigma )
\right)f^{(C)}(\sigma ,\tau )=0,$$

\noindent
which can be dealt with analytically.

Zero modes (modes with $\lambda (C)=0$) correspond to massless fluctuations on the
worldvolume theory of the D-string. It is natural to interpret these modes as Goldstone
bosons associated with the spontaneous breaking of translation, rotation and boost invariance
of the worldvolume theory of the brane. In the large $\sigma$ region where we can
use (for the D1$\perp$D3 system this is true for all $\sigma$)

$$ R(\sigma )\propto {1\over\sigma},$$

\noindent
we find that $f(\sigma ,\tau)$ satsifies

$$\left(-\partial_\tau^2+\partial_\sigma^2 -{\rho\over \sigma^2}
\right)f^{(C)}(\sigma ,\tau )=0 .$$

\noindent
Modes with $\lambda (C)<0$ correspond to modes with $\rho<0$. Making the ansatz
$f^{(C)}(\sigma ,\tau )=e^{i\omega\tau}g(\sigma )$, we find the following eigenvalue
problem

$$\omega^2g(\sigma )=\left(-\partial_\sigma^2 -{|\rho|\over \sigma^2} \right)g(\sigma )\equiv Hg(\sigma ).$$

\noindent
The eigenvalue problem for this quantum mechanical Hamiltonian has been studied previously; 
recent results include \cite{eig1},\cite{eig2}. We will review an argument given in \cite{eig1} 
which provides a simple illustration of the instability of the D1$\perp$D3 funnel for 
$\lambda (C)<0$. Consider

$$ {\langle\psi |H|\psi\rangle\over \langle\psi |\psi\rangle}=
{\int |\psi'(\sigma)|^2 d\sigma -|\rho|\int {|\psi(\sigma)|^2\over\sigma^2} d\sigma\over
\int |\psi(\sigma )|^2 d\sigma}.$$

\noindent
Set $\psi_\lambda (\sigma )=\sqrt{\lambda}\psi (\lambda \sigma)$ so that

$$ {\langle\psi_\lambda |H|\psi_\lambda\rangle\over \langle\psi_\lambda |\psi_\lambda\rangle}=
\lambda^2 {\int |\psi'(\sigma)|^2 d\sigma -|\rho|\int {|\psi(\sigma)|^2\over\sigma^2} d\sigma\over
\int |\psi(\sigma )|^2 d\sigma}.$$

\noindent
This demonstrates that if\cite{eig1}

$$|\rho | >{\int |\psi'(\sigma)|^2 d\sigma\over\int {|\psi(\sigma)|^2\over\sigma^2} d\sigma},$$

\noindent
then $H$ is unbounded from below indicating an infinite number of negative energy states. It is
possible to argue further that\cite{eig1} this will occur when $|\rho|>{1\over 4}$.

The existence of negative energy states (i.e. the existence of bound states
for the above Hamiltonian) implies that we have modes with $\omega^2<0$ and 
which are thus tachyonic. For a large enough negative eigenvalue we see that the D1$\perp$D3 
system has an infinite number of tachyonic modes. Thus, instabilities of the D1$\perp$D3 funnel
show up as negative eigenvalues $\lambda (C)$\footnote{The eigenvalues $\lambda (C)$ for the D1$\perp$D3
system are in fact all positive, as expected, since the D1$\perp$D3 system is supersymmetric and hence
obviously stable.}. 

These results do not immediately generalize to the D1$\perp$D5 and the D1$\perp$D7 funnels, 
because the bound states have their support at small $\sigma$ which is precisely the region 
in which the funnels differ. For example, the Hamiltonian relevant for the D1$\perp$D5 funnel, 
for small $\sigma$ is 

$$\omega^2g(\sigma )=\left(-\partial_\sigma^2 -{|\rho|\over \sigma^{2/3}} \right)g(\sigma )
\equiv Hg(\sigma ).$$

\noindent
Although this potential is not as ``steep" as for the D1$\perp$D3 system, we still expect bound
states (signalling tachyons) for large enough $|\rho |$. Modes with $\lambda (C)>0$ correspond to 
a repulsive potential and hence there can be no bound states.

Thus, in conclusion modes with $\lambda (C)<0$ need to be considered carefully as they may indeed
signal instabilities. 

\section{Results for the D1$\perp$D5 System}

The size of the matrix $\langle \{B,j\}|L|\{A,i\}\rangle$ grows
rapidly with $n$. For $n=1$ there are a total of 16 basis elements 
and the matrix $\langle \{B,j\}|L|\{A,i\}\rangle$ is a 128$\times $128 
dimensional matrix; for $n=2$ there are a total of 100 basis elements and the matrix
$\langle \{B,j\}|L|\{A,i\}\rangle$ is an 800$\times $800 dimensional matrix; for $n=3$,
$\langle \{B,j\}|L|\{A,i\}\rangle$ is a 3200$\times $3200 dimensional matrix.
In this section we will discuss the results obtained for $n=1,2$.

\subsection{$n=1$}

If we had been working on a commutative sphere, then modes with an angular
momentum less than or equal to 1 would be expressed in terms of the scalar or the $l=1$
partial waves. Here the scalar mode is replaced by the identity
matrix and the first partial waves by the $X^i$. We call fluctuations
that can be expressed in terms of the identity and $X^i$ the classical contribution.
Since the coordinates do not commute we also have contributions that are expressed
in terms of $\big[X^k,X^l\big]$. These have no analog in the case of a commutative 
geometry and hence we refer to these fluctuations as the quantum contribution.
At this order and at $n=1$, to obtain the full set of fluctuations, we have checked that there is 
no need to mix the new quantum degrees of freedom and the degrees of freedom that
would be present for a commutative sphere. However, we expect this to change once 
interactions are included and have verified that there is mixing for $n>1$. 
Further, looking at the eigenvalues of the fluctuations, we
see that the dynamics does not discriminate against the new degrees of freedom; they
have eigenvalues that are comparable to the eigenvalues of the degrees of freedom that
would be present for a commutative sphere. The spectrum of eigenvalues is

$$\matrix{
{\rm Eigenvalue} &{\rm Degeneracy} &{\rm Classical\,\, Contribution} 
&{\rm Quantum\,\, Contribution}\cr
0                &43               &8                                &35\cr
8                &44               &14                               &30\cr
16               &30               &25                               &5\cr
24               &10               &0                                &10\cr
48               &1                &1                                &0}$$

\noindent
Since there are no negative eigenvalues, we see that to the order we are
working, the D1/D5 funnel is stable.

The modes with eigenvalue equal to zero are giving us information about the 
moduli space (that is, the space of solutions) of the fuzzy sphere. Intuition from the
commutative case suggests that this moduli space is generated by performing translations
and $SO(1,5)$ transformations on the fivebrane worldvolume. Looking at the above results,
we find that there are 8 zero modes corresponding to translations of the $X^i$

$$\delta\Phi^i\propto 1.$$

\noindent
The $SO(1,5)$ transformations would include 5 boosts 
and 10 rotations (symmetries of $S^4$). However, 
we have a remaining 35 zero modes that are all excitations 
involving only the new quantum degrees of 
freedom $\big[X^i,X^j\big]$. It has already been argued 
\cite{sanjaye},\cite{ho} that the non-commutative $S^4$
can be viewed as a fuzzy $S^2$ fiber over an $S^4$ base. 
The symmetry of a commutative $S^2$ is $SO(3)$
and of a commutative $S^4$ is $SO(5)$. Assuming the symmetries 
acting on the base and fiber act independently
one might expect that the 10 rotations are replaced by 
$3\times 10=30$ rotations when we consider the 
non-commutative $S^4$. Thus, it seems natural to expect
$30$ rotations and $5$ boosts giving $35$ zero modes for the fuzzy funnel in complete 
agreement with what we have obtained numerically.
If this interpretation is correct, we see that the zero 
mode spectrum reflects, in a rather transparent 
way, the extra dimensions present in the fuzzy sphere geometries\cite{sanjaye},\cite{ho}.

\subsection{$n=2$}

In this case the generic fluctuation is a linear combination of the new quantum degrees of freedom and the 
degrees of freedom that would be present for a commutative sphere. The spectrum of eigenvalues is

$$\matrix{
{\rm Eigenvalue} &{\rm Degeneracy}\cr
0                &148\cr
8                &44\cr
16               &300\cr
24               &40\cr
32               &140\cr
40               &42\cr
48               &71\cr
72               &10\cr
80               &5}$$

\noindent
Again, the absence of negative eigenvalues implies that the solution is stable.

We again find 8 zero modes corresponding to translations of the $X^i$. The interpretation of the remaining 140
zero modes does not seem to be so simple. Assuming our interpretation for the zero modes given above is correct,
the fact that $140-5=135$ is not a multiple of 30 suggests that the symmetries acting on the base and fiber no
longer act independently. Clearly, clarifying the connection between the zero modes and symmetries of the 
non-commuative geometry in general remains an interesting open problem.

$$ $$

\noindent
{\bf Results for the D1$\perp$D7 System}

We will only consider the case $n=1$. There are 64 basis elements and the matrix 
$\langle \{B,j\}|L|\{A,i\}\rangle$ is a $512\times 512$ dimensional matrix. In this case, 
it is again possible to classify the small fluctuations as classical or quantum. 
Here the quantum contributions include both $\big[ X^i,X^j\big]$ and 

$$ X^{ijk}=X^{i}X^{j}X^{k}+X^{j}X^{k}X^{i}+X^{k}X^{i}X^{j}
-X^{i}X^{k}X^{j}-X^{j}X^{i}X^{k}-X^{k}X^{j}X^{i}.$$

\noindent
This is a direct consequence of the fact that the Young diagrams for
$SO(7)$ have three rows. The spectrum of eigenvalues is

$$\matrix{
{\rm Eigenvalue} &{\rm Degeneracy} &{\rm Classical\,\, Contribution} 
&{\rm Quantum\,\, Contribution}\cr
0                &113              &8                                &105\cr
8                &210              &0                                &210\cr
16               &62               &27                               &35\cr
24               &105              &28                               &77\cr
48               &21               &0                                &21\cr
72               &1                &1                                &0}$$

\noindent
The absence of negative eigenvalues suggests that this solution is perturbatively stable. There are 113 zero modes.
We expect 8 of these zero modes correspond to translations. The $SO(1,7)$ symmetry
of the worldvolume theory of the D7 branes includes
7 boosts. The remaining 98 zero modes are presumably associated with symmetries of the fuzzy $S^6$.

\section{Comparison to Supergravity}

In this section we study the linearized equations for the full Born-Infeld action. For the D3$\perp$D1 system, these
linearized equations correctly reproduce some of the modifications of the supergravity background generated by the D3
branes\cite{sugra},\cite{sugra2},\cite{const}. For the D5$\perp$D1 system a test-string in the background generated by $n$ D5-branes
does not agree with the linearized equation describing small fluctuations of the D1$\perp$D5 funnel\cite{const2}. 
An understanding of this discrepancy is required for a proper interpretation of the fuzzy funnel.

\subsection{Linearized Born-Infeld}

Following\cite{const2} we consider the fluctuation

$$\delta\Phi^6 (\sigma,\tau )=f(\sigma ,\tau ){\bf 1},$$

\noindent
with ${\bf 1}$ the $N\times N$ identity matrix. Plugging this into the D-string action
(in the background of the funnel $\Phi^i_0=RG^i/\sqrt{c}\lambda$) we obtain

\begin{eqnarray}
S=&-&NT_1\int d^2\sigma\sqrt{(1+R^{\prime 2})(1-\lambda^2\dot{f}^2)+\lambda^2 f^{\prime 2}}
\big[ 1+4R^4/(c\lambda^2)\big]\nonumber\\
=&-&NT_1\int d^2\sigma\left(H-{1\over 2}\lambda^2 H\dot{f}^2 +{1\over 2}f^{\prime 2}
\right)\qquad H(\sigma )=\left(1+{4R(\sigma )\over c\lambda^2}\right)^2 .
\nonumber
\end{eqnarray}

\noindent
Only the quadratic terms are kept as these determine the linearized equation of motion

$$ 0=(H\partial_\tau^2 -\partial_\sigma^2 )f(\tau,\sigma )\approx
\left(\left[ 1+{n^2\lambda^2\over 8\sigma^4}\right]\partial_\tau^2 -\partial_\sigma^2 
\right)f(\tau,\sigma ).$$

It is natural to ask about the dependence of this linearized equation of motion on the
specific matrix structure assumed for the fluctuation we consider. Towards this end, 
note that to evaluate the D-string action, we need to compute the terms quadratic in 
$\delta\Phi^6$ in the determinant

$$\det\left[\matrix{
-1 &0  &0 &0 &0 &0 &0 &\lambda\partial_\tau\delta\Phi^6\cr
 0 &1  &\lambda\partial_\sigma\Phi^1 &\lambda\partial_\sigma\Phi^2 &\lambda\partial_\sigma\Phi^3 
&\lambda\partial_\sigma\Phi^4 &\lambda\partial_\sigma\Phi^5 &\lambda\partial_\sigma\delta \Phi^6\cr
 0  &-\lambda\partial_\sigma\Phi^1 &1     &Q^{12} &Q^{13} &Q^{14} &Q^{15} &Q^{16}\cr
 0  &-\lambda\partial_\sigma\Phi^2 &Q^{21} &1      &Q^{23} &Q^{24} &Q^{25} &Q^{26}\cr
 0  &-\lambda\partial_\sigma\Phi^3 &Q^{31} &Q^{32} &1      &Q^{34} &Q^{35} &Q^{36}\cr
 0  &-\lambda\partial_\sigma\Phi^4 &Q^{41} &Q^{42} &Q^{43} &1      &Q^{45} &Q^{46}\cr
 0  &-\lambda\partial_\sigma\Phi^5 &Q^{51} &Q^{52} &Q^{53} &Q^{54} &1      &Q^{56}\cr
-\lambda\partial_\tau\delta\Phi^6 &-\lambda\partial_\sigma\delta\Phi^6
        &Q^{61} &Q^{62} &Q^{63} &Q^{64} &Q^{65} &1
}\right],$$

$$ Q^{ij}=i\lambda\big[\Phi^i,\Phi^j\big] .$$

\noindent
It is easy to see that to $O((\delta\Phi^6)^2)$, the $\sigma$ dependence of the coefficients of both $(\delta\dot{\Phi}^6)^2$ and 
$(\delta{\Phi}^{\prime 6})^2$ are unchanged. New terms of the form $(\delta\Phi^6)^2$ and $\delta\Phi^6\delta\Phi^{\prime 6}$
may be generated. This is enough to see that the $\sigma$ dependence of the $\partial_{\tau}^2\delta\Phi^6$ and
$\partial_{\sigma}^2\delta\Phi^6$ terms in the linearized equation of motion are independent of the
specific matrix structure assumed for the fluctuation we consider.

\subsection{Supergravity Analysis}

The results of the previous section should correspond to the fluctuation equation for a test D-string in some 
supergravity background. Assuming a diagonal background metric the test D-string action is\cite{const2}

$$S=-NT_1\int d^2\sigma \left(\sqrt{h}-{1\over 2}\lambda^2 h^{3/2}\dot{f}^2
+{1\over 2}\lambda^2\sqrt{h}f^{\prime 2}\right).$$

\noindent
For $n$ D5 branes

$$ ds^2 =h(\sigma )^{-{1\over 2}}\eta_{\mu\nu}dx^\mu dx^\nu + h(\sigma )^{-{1\over 2}}
(d\sigma^2 +\sigma^2 d\Omega_3^2),\qquad h(\sigma )=1+{L^2\over\sigma^2} ,$$

\noindent
the resulting linearized equation of motion for $f$ can be written as\cite{const2}

$$(h^2\partial_\tau^2-\partial_{\tilde{\sigma}}^2)f(\tilde{\sigma},\tau)=0,\qquad
\tilde{\sigma}^2=\sigma^2+L^2 .$$

\noindent
This does not match the result following from expansion of the Born-Infeld action, as first observed in \cite{const2}.
Further, as argued in the previous section, this mismatch is independent of the specific
matrix structure assumed for the fluctuation of the fuzzy funnel description. This disagreement is puzzling
as both the D-string and test-string descriptions should be valid for large $\sigma$. The disagreement can be traced
back to the fact that all three funnel solutions have the same large $\sigma$ behaviour, so that the fluctuations
from all three funnels look similar to the D3$\perp$D1 system.

\subsection{Newtonian Approximation} 

In this section we will build a Newtonian gravity description of the funnel. Our analysis provides
a toy model description of the full dynamics that illustrates some subtle points, explaining
the mismatch between the supergravity and linearized Born-Infeld descriptions.

To compare the Newtonian and supergravity descriptions, recall the well known fact that for a particle moving
with a small velocity, in a region of spacetime well approximated by a flat metric,
the geodesic equation becomes (we take $\tau$ to be an affine parametrization)

$$ {d^2x^i\over d\tau^2}=\Gamma^i{}_{\alpha\beta}{dx^\alpha\over d\tau}{dx^\beta\over d\tau}\approx
{1\over 2}{dg_{00}\over dx^i},$$

\noindent
so that we can compare $g_{00}$ to the gravitational potential computed in the Newtonian theory.

A segment of the funnel stretching from $\sigma$ to $\sigma +d\sigma$ has a mass ($d=2$ for the D1$\perp$D3 system
and $d=4$ for the D1$\perp$D5 system)

$$ dm=T_{d+1}\sqrt{1+R'(\sigma)}R^d(\sigma )\Omega_d d\sigma .$$

\noindent
In 9 spatial dimensions, the potential of a point mass drops off as $1/r^{7}$. Thus,
the potential set up by the funnel is

$$\phi =\int {T_{d+1}\sqrt{1+R'(\sigma)}R^d(\sigma )\Omega_d\over r^7}d\sigma ,$$

\noindent
where $r$ is the distance from the observation point ($\vec{x}_{obs}$) to the segment of the funnel 
stretching from $\sigma$ to $\sigma +d\sigma$

$$ r=|\vec{x}_{obs}-\vec{x}_{\rm funnel\,\, segment}|,$$

$$\vec{x}_{\rm funnel\,\, segment}=(x^0,\sigma,R(\sigma )\vec{\Omega},0,\vec{0}),$$

$$\vec{x}_{obs}=(x^0,\sigma_o,\vec{0},x^{d+2}_o,\vec{0}).$$

\noindent
In what follows we will discuss only the D1$\perp $D5 system. We have checked that exactly
the same behaviour is exhibited by the D1$\perp$D3 system. Consider the integrand in the expression
for $\phi$. We will study this integrand as a function of $\sigma$ for different values of $x^6_o$.
The behaviour we will discuss below is qualitatively independent of the details of
$R(\sigma )$. For this reason we will take $R(\sigma )={1\over\sigma}$ to illustrate our arguments.
We will set $\sigma_o =2$. For $x^6_o=2$, the integrand is shown in figure 1.

\begin{center}
\begin{figure}[h]{\psfig{file=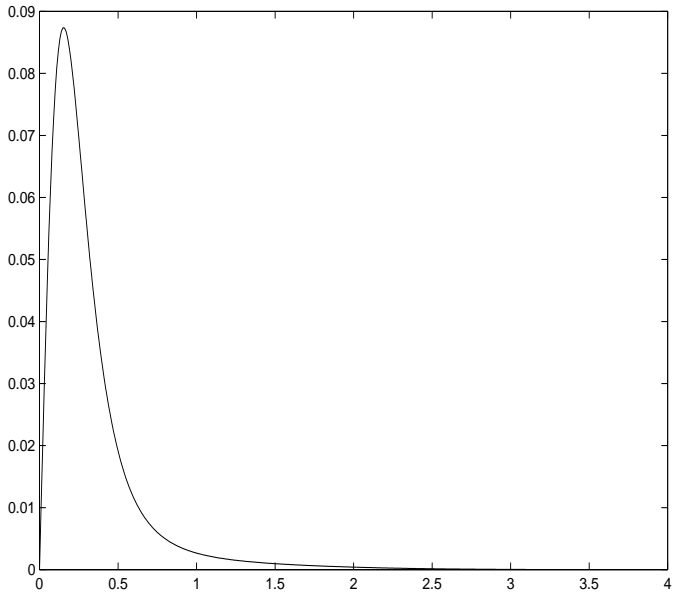,width=9cm,height=8cm}
\caption{The integrand for the funnel potential for $x^6_o =2$.}}
\end{figure}
\end{center}

\noindent
The integral is clearly dominated by the contribution coming from $\sigma\approx 0$, that is, the
potential at this observation point is dominated by the contribution from the D5 brane. 
Decreasing $x_o^6=0.8$, the integrand behaviour is shown in figure 2.

\pagebreak

\begin{center}
\begin{figure}[h]{\psfig{file=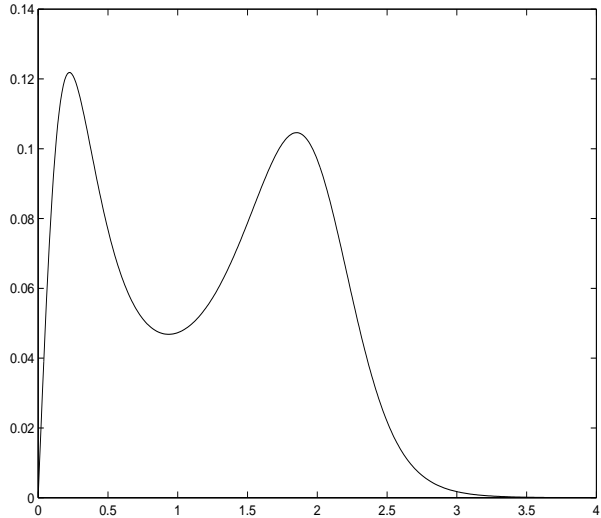,width=8cm,height=7cm}
 \caption{The integrand for the funnel potential for $x^6_o =0.8$.}}
\end{figure}
\end{center}

\noindent
We now have two significant contributions - a D5 brane contribution at $\sigma\approx 0$ and 
a contribution from $\sigma\approx 2$ corresponding to that portion of the funnel closest to 
the observer, which for large $\sigma_o$ looks like a D-string contribution. This behaviour
is easy to understand on physical grounds - as we move closer
to the funnel, the local geometry has a bigger effect. Decreasing
$x_o^6=0.1$, the integrand behaviour is shown in figure 3.

\begin{center}
\begin{figure}[h]{\psfig{file=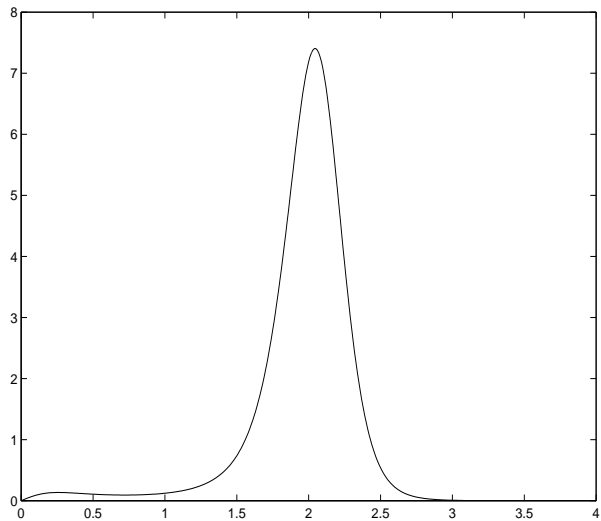,width=8cm,height=7cm}
 \caption{The integrand for the funnel potential for $x^6_o =0.1$.}}
\end{figure}
\end{center}

\noindent
The integrand is now dominated by a D-string contribution - the contribution from the D5 brane
at $\sigma\approx 0$ is negligible. Assuming the funnel reproduces the gravitational potential
in this third region, we have a convincing explanation of why the fluctuations described by
the funnel are the same for both the D3$\perp$D1 and the D5$\perp$D1 systems: only the geometry 
of the funnel for large $\sigma$ is relevant, and this is the same for both funnels. 
A comment is in order. In our plots the integrand peaks at $\sigma =2=\sigma_o$. We chose this
value of $\sigma_o$ to produce plots which clearly illustrate our point. In the computation
of small fluctuations which can be compared to a test string one would take $\sigma_o>>1$ large.

We have not yet discussed what we mean by a ``small fluctuation". We have not
pursued a suitable definition of what a ``small" fluctuation in this Newtonian toy model
is. Rather, we have been content to show that there is a choice for the scaling of the 
fluctuation that (i) ensures that the funnel predictions are reproduced and (ii) for
which the fluctuations are small. By reproducing the funnel predictions, we mean we
find a potential with a ${1\over\sigma_o^4}$ dependence. Making the ansatz

$$ x^{d+1}\propto\sigma_o^{-{1\over n}},$$

\noindent
we find that with $n={5\over 13}$ for the D3$\perp$D1 system and with $n={7\over 13}$
for the D5$\perp$D1 system, the a ${1\over\sigma_o^4}$ dependence for the potential
is obtained. To obtain these values we have performed a steepest descent evaluation
of the potential assuming that $\sigma_o$ is large. The integrand has a peak at 
$\sigma\approx\sigma_o$. It is pleasing that the fluctuations are indeed small in the large
$\sigma_o$ region we focus on, showing some degree of self consistency.

To summarize, our toy model shows that the background corresponding to a stack of $n$
D5 branes might not be the appropriate background against which to compare the funnel
results. Further, we have shown that, for a very specific assumption about the $\sigma_o$
dependence of the fluctuations, the gravitational results and the funnel results are 
consistent. It would be nice to perform a more complete analysis in supergravity, including
a careful independent determination of the scaling of $x^{d+1}$ with $\sigma_o$, to
confirm that the two are indeed consistent. 

\section{Summary}

In this article we have considered the non-commutative fuzzy funnel geometries that provide
a description of a D-string ending on a D3, D5 or D7 brane. In particular, we have focused
on small fluctuations about the leading funnel geometry. A numerical treatment of these
small fluctuations has been given. The connection between zero modes and symmetries of
fuzzy sphere geometries has been discussed. This connection is particularly transparent
for the D1$\perp$D5 system with $n=1$. The comparison of the linearized fluctuations 
of the full Born-Infeld action and the description of these fluctuations in supergravity
has been considered in the framework of a Newtonian toy model. An explanation of 
previous disagreements between the two descriptions has been suggested. 

There are a number of ways in which this work can be extended. The connection between zero
modes and symmetries of the fuzzy sphere deserves to be studied further. This will involve a
careful study of the geometry of the fuzzy sphere. The results of \cite{san2} seem to provide
a possible approach for this study. Another interesting direction which could 
be followed would involve improving upon our Newtonian toy model of the funnel. The construction
of a supergravity model of the funnel as well as a demonstration that the D-string description
agrees with the supergravity description would be satisfying.

$$ $$

\noindent
{\it Acknowledgements:} We would like to thank Sanjaye Ramgoolam for very pleasant and helpful
discussions. We also thank Jeff Murugan and Sanjaye Ramgoolam for comments on the manuscript.   
This work is supported by NRF grant number Gun 2047219.

\end{document}